\documentclass[aps,prd,nofootinbib,superscriptaddress]{revtex4-2}
\usepackage{amsfonts}
\usepackage{amsmath}
\usepackage{amsthm}
\usepackage{amssymb}
\usepackage{mathtools}
\usepackage{enumitem}
\usepackage{textcomp}
\usepackage{gensymb}
\usepackage{hyperref}
\usepackage{booktabs}

\usepackage{bm}

\usepackage{xcolor}
\usepackage{dsfont}

\usepackage{lmodern}

\usepackage{graphicx}
\graphicspath{ {figures/} }
\usepackage{tipa}

\usepackage{tikz-cd} 

\DeclareMathOperator{\Tr}{Tr}
\newcommand{\Der}{\mathrm{D}}
\newcommand{\der}{\mathrm{d}}

\newcommand{\tetrad}{\text{\textschwa}}

\begin{document}
\title{
	{Nontrivial constitutive laws and unified structures in constrained BF theory}
}
\author{Priidik Gallagher}
\email{priidik.gallagher@ut.ee}
\affiliation{Laboratory of Theoretical Physics, Institute of Physics, University of Tartu, W. Ostwaldi 1, 50411 Tartu, Estonia}

\begin{abstract}
	Does a nontrivial gravitational excitation require a modified internal gauge theory constitutive law? As there is no canonical mapping between differential forms valued in distinct Lie algebras, the answer is negative, and entirely dependent on the specific unification scheme. A structural formulation in BF theory in terms of a constitutive diagram between the excitations of different interaction sectors is provided, alongside a discussion of the structure of the broken phase. As a nontrivial option, a ``spontaneous'' breaking into the physical constraint is attempted, however it is shown that basic B-potentials alone would not be viable. A heuristic discussion of internal gauge theory and gravity is provided, and by conflating the observer's internal and external state with the spacetime tangent structure, it is argued that there is a simple geometric obstruction to a nontrivially unified phase. A more \textit{ad hoc} treatment of gauge theory and gravitational structure remains as the clear path forward, while observer, signal and causal considerations would suggest studying alternative backgrounds to the manifold topology.
\end{abstract}

\maketitle

\section{Introduction}

Although significant effort has gone into developing a variety of premises for field theory, gauge-gravity unification remains an elusive topic. Gauge interactions have been straightforwardly geometrized in terms of connections on principal bundles~\cite{Mielke:2017nwt}, while fermionic matter fields are described in spin geometry~\cite{Lawson_Michelsohn:1989}. From a different side, advances in topological field theories have long proven their relevance in standard physics~\cite{Birmingham:1991}, but the interrelation between gauge theory and gravity is rather surprising, such as 3-dimensional General Relativity appearing as a Chern-Simons theory~\cite{Witten:1988,Carlip:1998uc}. Among various topological field theories, BF theory appears as a particularly generic dynamical premise for gauge theory, as it is simply the differential form top rank completion $\Tr(B\wedge F)$ of a curvature term $F$ by an additional field $B$. The physical content becomes contingent on how the $B$-field is constrained to its physical value, which in premetric terms~\cite{Hehl_Obukhov:2003} is the constitutive law of the gauge theory, or in gravitational terms the simplicity constraint~\cite{Livine:2007ya,Livine:2025}. It is one of the main distinctions between internal gauge theory and gravity~\cite{Weatherall2016-WEAFBY,March_Weatherall:2024,Gomes:2024}, and provides an impression of symmetry breaking, albeit in a function space, so it seems plausible that a gauge-gravity unified phase might be of topological type. In terms of Einstein-Cartan gravity and Yang-Mills theory, there seems to be a particular physical fixing of $B_\text{gravity}=\frac{1}{2\kappa}*(e^a\wedge e^b)$ vs.~$B_\text{gauge}=\frac{1}{2g^2}*F$, and it is the extent and interrelation of these choices that is the question of study here.

The purpose of this paper is several-fold. The main objective is to study whether a non-standard constitutive law of one component necessitates a modification in the the constitutive law of the other. The ``Khronon'' Lorentz gauge theory~\cite{Zlosnik:2018qvg} is a particular example of nontrivial gravitational constitutive structure, compared to the standard General-Relativity or Einstein-Cartan hypersurface basis. The answer itself is negative, due to the fact that there is no canonical structure which would implicate the opposite, and this is formalized by introducing and studying the constitutive diagram between internal and external excitations as a new, separate concept. In essence, this is a different description of the fact that gravity and internal gauge theory describe different sectors of geometry. The implication is that the specific emergence of gauge theory and gravity is rather \textit{ad hoc} on a per-theory basis.

Alongside this, a general description of the basic internal-external constitutive breaking is given, including by considering an inter-gauge rotation. Here, ``internal'' refers collectively to Maxwell electromagnetism and Yang-Mills theory, as the principal bundle and Lagrangian structure is similar and internal symmetries, as commonly used, are automorphisms of the principal bundles, while ``external'' refers to gravity, in the sense of external symmetries which are diffeomorphisms of spacetime. Although the broken structure, that is, the phase where the structure groups are separated into a direct product and the Lagrangians separate into a sum, is straightforward, the simultaneous treatment does not appear to be extensive in literature, with the focus instead on generating the split, rather than studying the broken phase in BF terms. The crucial issue in constrained theory is that resolving the constraints also resolves the theory to its final form, so if the constraint structure is unique, no physically substantial difference has been actually achieved. As a previously untested option, a ``spontaneous'' breaking into the constitutive law is attempted in terms of a polynomial potential, a generalized quartic potential and a general $V(B)$ potential. That is, the case when the equations of motion of the $B$-field admit multiple, distinct solutions is studied, of which one is the usual constitutive law $B=\kappa(A)$, while other minima are considered as non-physical. This introduces additional freedom or initial conditions into the theory, which must be resolved (broken into); in more traditional terms, this is a matter of resolving a degenerate set of equations.

This requires a slight generalization of the notion of symmetry breaking of dynamical fields, on the basis of an equation having multiple solutions; cf.~the vacuum manifold of the Higgs mechanism~\cite{Beekman:2019pmi,George:2013,Elliot:2021,Gaiotto:2014kfa}. Spontaneous symmetry breaking proper is, as usual, the loss of (Lie group) symmetries in the states or the solutions compared to the Lagrangian, but a characteristic of this phenomenon is a nontrivial vacuum manifold, so the symmetry breaking field can obtain one of a multitude of possible values. It is this degeneracy in solutions that is taken as the basis for generalization for differential forms and, although it does not bear a clear interpretation in terms of (Lie group) symmetries. So, when a degenerate potential for the $B$-field would be devised, an additional choice is required, which would mimic a ``spontaneous'' choice of the constitutive law of the theory. However, it is shown that spontaneous breaking via a pure $B$-potential is not feasible due to back-reaction to the inhomogeneous equation, requiring a different degenerate constraint structure for a meaningful path forward. Plebanski theory~\cite{Plebanski:1977,Krasnov:2010olp,Krasnov:2009b} is a particular example of such nontrivial constraint structure, as it also permits several classes of solutions for the $B$-field.

Premetric theory aims to axiomatize and develop physical theories as extensively as possible without referring to the spacetime metric. The case of both electromagnetism and gravity has been extensively studied~\cite{Hehl_Obukhov:2003,Hohmann:2018,Itin:2018dru,Obukhov:2019oar,Puetzfeld:2019,Koivisto:2021}, but there does not appear to be extensive literature relating these ideas to topological field theory, so another aim of this paper is to further clarify this relation. As topological field theory does not make use of the metric, deriving dynamics from it~\cite{Floreanini:1990cf,Akama:1990ur,Antonsen:1994fka,Anselmi:1994fs,Anselmi:1997,Cattaneo:1997eh,Oda:2016tdr,Cattaneo:2023cnt} can be seen as an execution of premetric ideas, and the basic statement for BF theory is almost trivial: the separate $B$-field is the excitation in the inhomogeneous equation $\Der B=J$, and is fixed to $*F$ or $*(e^a\wedge e^b)$ through some dynamical realization of the constitutive law, while the remainder of premetric axiomatics is absorbed into the principal bundle structure. At the same time, this breaking appears to prefer particular values, of (Maxwell-)Yang-Mills or Einstein-Cartan type. A heuristic explanation for this will be discussed in the sense of internal gauge theory vs.~gravity throughout, controlling the geometry of either some imposed, additional principal bundle (in essence, its connection), or the objects that are canonically induced or directly related to the underlying spacetime (instead, the soldering form or tetrad and frame bundle connection).

Only the trivially broken phase will be considered, where the internal principal bundle $P_\text{int}$ and the frame bundle $F(M)$ are not intermixed, $P_\text{tot}=P_\text{int}\oplus F(M)$. If internal gauge theory and gravity were unified, it is possible that this geometry is different, but the specifics are generally model-dependent. This is sufficient to study the interrelation of the $B$-values in the broken phase, and similarly it is conformant to the Coleman-Mandula theorem~\cite{Coleman_Mandula:1967}. Generally, this implies that either more or different structure is required for a gauge-gravity unified phase, and there are many approaches to resolve this issue~\cite{Krasnov:2018}. In turn, a simple argument is provided for a geometry change in the unified phase, in that the manifold tangent structure is incompatible with the nontrivial observer state in the unified phase when trying to immediately construct spacetime.

The paper is structured as follows. Section~\ref{sec:spacetime_matter} provides a brief review of gauge theory and gravity in terms of principal bundles, and a heuristic argument is made for identifying the gravitational excitation with the hypersurface basis. In~\ref{sec:synchronization}, a simple geometric obstruction is described for a nontrivial unified phase, in terms of constructing spacetime as a field of observers. The remainder considers the structure of BF theory. Section~\ref{sec:BF_theory} provides the basic structure how the BF theory sectors separate. Spontaneous breaking into the constitutive law is considered in section~\ref{sec:duality}. Finally, section~\ref{sec:constitutive_laws} introduces the constitutive diagram, and provides an example of introducing a nontrivial tetrad structure to Plebanski theory.

\section{Signal geometry}
\label{sec:signal_geometry}

\subsection{Spacetime and matter}
\label{sec:spacetime_matter}

To motivate the Lagrangian-level internal-external questions later on, let us first consider the geometry of spacetime and matter. The discussion will be classical, with the purpose to emphasize how the soldering of the tangent bundle relates to a local notion of external space, thus to gravitation. Internal space is inhabited by matter, described by fiberwise directions in principal and associated bundles, and attached to external space, constituted by the spacetime manifold itself. Put another way, the issue faced is that the spacetime manifold already provides the frame bundle structure, which is a natural arena for the theory of gravity. Arguably, a truly unified phase might not have such distinction, which is the reason for introducing this terminology: gravitational degrees of freedom are to be immediately related to objects of spacetime geometry, such as the metric or the tetrad and the connection, rather than as a separate (graviton or otherwise) field present in a background spacetime.

Commonly, spacetime is built as the set of events $M$, sufficiently bound together by adequate topological axioms and requirements on differentiability, producing a four-dimensional smooth manifold. Frequent requirements (with some useful implications) include $M$ being Hausdorff-separable (vs.~topologically degenerate points), path-connected (vs.~various (worm)holes), paracompact (partition of unity) and oriented (vs.~lacking globally consistent tempora-spatial directions); their specific appearance, interrelation\footnote{For example, Hausdorff-separability and the Lorentz metric imply paracompactness~\cite{Geroch:1968zm}. But if introducing the metric is to be avoided, as in the pre-metric programme, this becomes a separate requirement.} and validity is nontrivial~\cite{Penrose:1972,Papadopoulos:2018uid,Wu:2023rmq}. Relativistic causality is introduced by a Lorentzian metric field $g$, which can be understood as a particular section of the bundle of symmetric and nondegenerate rank 2 tensors,
\begin{equation}
	g=g_{\mu\nu}\der x^\mu\otimes \der x^\nu\in\Gamma(T^*M\otimes T^*M).
\end{equation}
This introduces a light cone causal structure and observers as unit timelike vector fields~\cite{Gielen:2013, Hohmann:2016}. Comparing tangent spaces at different points requires the (linear) connection $\Gamma^\rho_{\mu\nu}$, equivalently the spin connection 1-form $\omega_{ab}$. For the purposes of General Relativity, the connection can be restricted to Levi-Civita. The trajectories are weighed by the Einstein-Hilbert action, with various equivalent expressions,
\begin{equation}
	S_\text{GR}=\frac{1}{2\kappa}\int \der^4x\sqrt{-g}R[\Gamma]\sim\frac{1}{2\kappa}\int *(e^a\wedge e^b)\wedge R_{ab}[\omega],
\end{equation}
which sets up the dynamical premise. There is great variety in the exact details, whether Einstein-Cartan, Palatini, metric-affine or otherwise~\cite{Krasnov:2020lku}, but altogether, this is a textbook construction~\cite{Misner:1973prb, Hawking:1973uf, Wald:1984rg}, and the structure is readily understood through the theory of pseudo-Riemannian manifolds and connections. Our focus will be on the $*(e^a\wedge e^b)$ term, viz.~the excitation in premetric terminology, which is not widely used for gravity, with the coframe $e^a$ and the Lorentz connection $\omega_{ab}$ as the basic variables.

We will use differential forms,
\begin{equation}\label{eq:dif_form_components}
	\alpha=\frac{1}{k!}\alpha_{\mu_1\ldots \mu_k}\der x^{\mu_1}\wedge\ldots\wedge\der x^{\mu_n}\in\Omega^k(M),
\end{equation}
as they are geometrically natural, and will particularly support the Lagrangian considerations. Although, by construction, differential forms and tensor components provide equivalent descriptions to field theory on curved spacetime, the latter does obfuscate some structure (specifically, the gravitational excitation) which is easily visible in the former. So, the (Lorentzian) coframe 1-form is $e^a=e^a{}_\mu\der x^\mu$ with the dual frame vector field $\tetrad_a=\tetrad_a{}^\mu\partial_\mu$, such that $e^a(\tetrad_b)=\delta^a_b$. It is equivalent to the metric in that
\begin{equation}
	g_{\mu\nu}=\eta_{ab}e^a{}_\mu e^b{}_\nu,
\end{equation}
where $\eta$ is an inner product on the tangent spaces. Of relativistic importance is the $SO(1,3)$ Lorentzian $\eta_{ij}=\mathrm{diag}(-,+,+,+)$, but $\eta$ can be relaxed to consider various other metrics. For example, consider the Hermitian scalar product on $\mathbb{C}^n$,
\begin{equation}
	\langle u,v\rangle=u^\dagger v=\eta_{ij}\bar{u}^i v^j.
\end{equation}
Here, in component form trivially $\eta=\mathrm{diag}(+,+,\ldots,+)$. The unitary group would appear as the set of invariance transformations of the scalar product,
\begin{equation}
	U(n)=\{g\in GL(n,\mathbb{C})\vert \langle g u, g v\rangle=\langle u,v\rangle\ \forall u,v\in\mathbb{C}^n\}.
\end{equation}
We consider the Lorentz (spin) connection 1-form $\omega_{ab}$ as a separate independent object, understood as the principal connection on the frame bundle. In Levi-Civita form, $\omega_{ab}$ would not be independent~\cite{Rodrigues:2007}, and instead
\begin{equation}
	\omega^{ab}=\frac{1}{2}(\tetrad^b\lrcorner\der e^a - \tetrad^a\lrcorner\der e^b + \tetrad^a\lrcorner(\tetrad^b\lrcorner\der e_i)e^i),
\end{equation}
where $\lrcorner$ denotes the interior product. The Palatini formalism resolves to this result dynamically. Altogether, this selection appears as one natural adjustment of gravitation to the premise of Yang-Mills theory, but it is hardly unique.

Of practical interest, note the Hodge star operator
\begin{equation}
	\begin{aligned}
		*:\Omega^k&\to\Omega^{n-k}\\
		\alpha&\mapsto *\alpha=\frac{\sqrt{\vert g\vert}}{k!(n-k)!}\alpha^{i_1\ldots i_k}\epsilon_{i_1\ldots i_k i_{k+1}\ldots i_n}\der x^{i_{k+1}}\wedge\ldots\wedge x^{i_n},
	\end{aligned}
\end{equation}
which provides the Hodge inner product on the space of differential forms,
\begin{equation}
	\begin{aligned}
		\langle\cdot,\cdot\rangle:\Omega^k\times\Omega^k&\to\mathbb{R}\\
		(\alpha,\beta)&\mapsto\int\alpha\wedge*\beta.
	\end{aligned}
\end{equation}
Lagrangian-wise, the Hodge star provides a crucial distinction from purely topological theory.

The differential geometry of principal bundles provides a similarly clean description for all gauge interactions, see~\cite{Hamilton:2017gbn,Nakahara:2003} and the many references therein. Then, the interaction is mediated by a gauge field $A$, given as a connection on a principal bundle $\pi:P\to M$ with structure group $G$, the gauge group. In practice, $A$ can be understood as a Lie algebra $\mathfrak{g}=T_e G$ valued 1-form on $M$, when suitably pulled back by a (global) gauge $s:M\to P$. This permits exterior calculus, e.g.~gauge transformations are
\begin{equation}
	A\to g A g^{-1} + g\der g^{-1}.
\end{equation}
The curvature $F$ is a particularly important object, given by e.g.~Cartan's structure equation
\begin{equation}
	F=\der A + A\wedge A.
\end{equation}
For gravity, also $F\equiv R$ for common notation. Overall, this description is geometrically natural, and standard phenomenology follows directly when given any specific Lagrangian. So, the structure group $U(1)$ yields electromagnetism, while non-Abelian groups provide Yang-Mills theory. In particular, note that the tangent orthonormal frame bundle $F_{SO}(M)$ is an $SO(m,n)$ principal bundle, compatible with the inner product $\eta(\cdot,\cdot)$ on the tangent spaces. In standard gravitational terms, the connection would be defined on the frame bundle, and this can be taken as a requirement for the connection variables for any would-be theory of gravity of Palatini type. But as emphasized earlier, this is hardly unique~\cite{Krasnov:2020lku}. For the frame bundle, principal bundle constructions apply. The principal Lorentz connection $\omega$ is built as in any gauge theory: see~\cite{Ferreiro_Munoz:2008,Capriotti:2014,Capriotti:2018,Rey:2020} and~\cite{Michor:2008} ch.~25.6~in particular. Of course, any distinction of gravitational, spacetime or otherwise in only meaningful in relation to specific dynamics and associated structures (spinors, covariant derivatives, etc.). As an example, it is straightforward to define extra matter content in the form of an (internal) $SO(1,3)$ gauge theory with a Yang-Mills Lagrangian\footnote{Which would be a non-compact gauge theory, with many difficulties~\cite{Cahill:1983,Margolin:1990wt,Margolin:1992rg,Alexander:2023}.}.

For terminology, external space can be globally understood as the base manifold $M$ itself, locally corresponding to an open neighborhood $p\in U\subset M$ charted by $\varphi:U\to\mathbb{R}^n$, and ultralocally the tangent space $T_p M$ at $p\in M$; in principle, the full tangent bundle itself can also be referred to external space, as $M$ determines $TM$ uniquely\footnote{The opposite appears true from the definition of $\pi:TM\to M$, which requires a prescripted $M$. However, if $M$ is parallelizable, an ambiguity could be introduced, in the sense that $M$ is diffeomorphic to any other global section. In a sense, this is akin to ``integrating'' $TM$ to $M$, and necessitates the introduction of a connection.}. The remainder is ``internal'' geometry, which can be understood as another fiber bundle $\pi:E\to M$, that is the totality of all fields at any given points, and thus has various subbundles. More immediately, the interest is in a principal $G$-bundle $P$ of all interactions, and its associated vector bundle $P\times_\rho V$ for matter fields. External symmetries are spacetime diffeomorphisms $f:M\to M$, internal symmetries on the gauge fields are principal bundle automorphisms $\varphi:P\to P$, which induce transformations on $P\times_\rho V$ through the representation $\rho$. It seems natural to also refer to the Lagrangian theories themselves using similar terminology, dependent on the geometry the objects are defined in. As such, theories of connections on separate principal bundles (separate from $F(M)$) would be collectively termed as internal gauge theories. However, any precise definition would also be dependent on the specific Lagrangian dynamics, which in BF theory terms, is contingent on the values the $B$-field is fixed to: we will return to this point shortly. For a precise summary of internal and external symmetries in terms of Lie algebroids~\cite{Mackenzie:2005,Catren:2014vza}, also consider the Atiyah exact sequence
\begin{equation}
	0\to P\times_G\mathfrak{g}\to TP/G\to TM\to0,
\end{equation}
with the adjoint bundle $P\times_G\mathfrak{g}$; we will not be required to go as in-depth. Instead, a quandary that we'd like to consider is that at a single point $p\in E$, provided only the tangent space $T_p E$, there must be a designation on which directions are external and which are internal, as if the observer somehow has global understanding of geometry. However, the specific capacity to split fiber-wise and base manifold-wise directions in the tangent space as horizontal and vertical
\begin{equation}\label{eq:Ehresmann_pointwise}
	T_pP=V_p\oplus H_p
\end{equation}
is part of the definition of the Ehresmann connection; specifically, it would introduce a distribution of right-invariant horizontal tangent spaces.

The purpose is to structurally separate the directions tangent to the base manifold from the remainder directions tangent to fibers, which are vertical to the fibers. The tangent bundle $TM$ is tautologically tangent to spacetime $M$, so any arbitrary vector field points can be understood to point in external directions\footnote{This provides another perspective to the nontrivial combination of notions $e^a\equiv\Der\tau^a$ in the gravitational theory of~\cite{Zlosnik:2018qvg}. Then, a vector field, or internal Lorentz-valued scalar, the ``Khronon'' field $\tau^a$, provides a measure of time.}. Soldering provides a formalization that a vector bundle $VM$ is tangent to the base manifold in terms of intrinsic geometry. Often, the coframe $e^a$ is simply termed the solder form, or the tautological form, but the logical purpose is to emphasize how the tangent bundle is indeed tangent to the base manifold. Following~\cite{Helein:2009,Minguzzi:2014,Catren:2014vza,Vey:2014rua}, let
\begin{equation}
	\varphi:TM\to VM
\end{equation}
be an isomorphism, i.e.~a section $\varphi\in \Gamma(VM\otimes_M T^*M)$ such that each $\varphi\vert_{p\in M}$ is an isomorphism. Equivalently, $\varphi$ is a 1-form with values in $VM$, the solder form. Clearly, the coframe $e^a$ is one such example, and a suitable object to model the geometry of spacetime.

A few comments to add. While the soldering of $\mathbb{R}^{1,3}$ onto $M$ is indeed unique to gravity~\cite{Catren:2014vza}, the gauge 1-forms $A$ can still be transformed into a collection of usual vector fields $g^{\mu\nu}A^I_\mu\partial_\nu\sim\sharp s^*A$ in terms of musical isomorphisms, indistinguishable from external directions apart from the internal gauge transformations,
\begin{equation}
	A^I_\mu\to A^I_\mu + \frac{1}{g}\partial_\mu\alpha^a + f^{IJK} A^J_\mu\alpha^K.
\end{equation}
So, internal directions are only meaningfully determined by their nontrivial interrelation. Secondly, soldering \emph{per se} requires a $VM\cong TM$ bundle, so $\dim VM=\dim TM$, but it is entirely reasonable to also interpret fibers and manifolds $F$ of other dimensions to also touch the base manifold in a tangent sense. This is straightforward to formalize, by denoting points of contact $p\in M$, $q\in F$ and identifying $T_pM\subseteq T_qN$ or $T_pM\supseteq T_qN$ as vector spaces; the result would not generally be a manifold as the points of contact may have a nontrivial neighborhood. See~\cite{Helein:2009} for a more rigorous discussion in terms of soldering.

The argument is that gravity, sprung from this differential geometric premise, is inherently a theory of the geometry of spacetime, so it necessarily relies on the notion of external space provided by the (co)tangent space. External directions are synonymous with spacetime directions, and this term is mostly meaningful from the viewpoint of an observer and in a gauge-gravity unification setting. Note that so far, there is no argument for what the structure group of gravity should be, although the most obvious physical candidates are the Lorentz group $SO(1,3)$, the (metric-affine) general linear $GL(4)$, the Poincaré group $ISO(1,3)$, or similar, but rather the discussion is what sectors of geometry the various interactions should even inhabit. Let it be emphasized that there is no unique and strictly necessary argument where the gravitational degrees of freedom must necessarily be located --- for example, it is entirely possible to consider pure connection models of gravity~\cite{Krasnov:2011,Delfino:2012zy,Alexander:2023}. Similarly, the distinction itself between gravity and gauge theory is a very classical question, very famously dating to Weyl, and still does not yet possess an unanimous resolution. Recently, the (dis)analogies have been considered in philosophical discourse~\cite{Weatherall2016-WEAFBY,March_Weatherall:2024,Gomes:2024}, while physically it is a motivation for various gauge approaches to gravity~\cite{Gronwald:1995em,Weinstein1999-WEIGAG,Catren:2014vza,Blagojevic:2013,Obukhov:2022khx}. Here, with the aim to match the setup of General Relativity, we follow the Palatini formulation in terms of $e^a$ and $\omega_{ab}$. So, let us consider some more rephrasings of basic concepts in terms of premetric theory. Briefly, the excitation $B$ is an $n-2$ form motivated from dynamics as a separate concept to field strength, easy to define as the complement to curvature in the Lagrangian $n$-form,
\begin{equation}\label{eq:excitation_lagrangian}
	L\sim \Tr(B\wedge F),
\end{equation}
but originates from the inhomogeneous Maxwell or Yang-mills equation
\begin{equation}\label{eq:charge_conservation}
	\Der B=J,
\end{equation}
when derived purely from charge and flux conservation. The particular physically relevant local and linear constitutive law is
\begin{equation}\label{eq:constitutive_gauge}
	B_\text{gauge}=\frac{1}{2g^2}*F,
\end{equation}
relating $B$ to the Hodge dual of the field strength $F$; see~\cite{Hehl_Obukhov:2003,Hehl:2004zz} for details on electromagnetism. Gravity appears to instead require
\begin{equation}\label{eq:constitutive_gravity}
	B_\text{gravity}=\frac{1}{2\kappa}*(e^a\wedge e^b).
\end{equation}
This reproduces the usual Einstein-Cartan formalism, and allows the connection to be entirely integrated out, purely leaving the (co)frame, viz.~the metric, as usual in Palatini formalism. 

The excitation was preemptively labeled $B$, so that the premetric Lagrangian~\eqref{eq:excitation_lagrangian} would immediately relate to BF theory. The distinction is only that in BF theory the excitation is taken as a separate dynamical field, and without any additional potential it would produce a purely topological theory (viz.~fixing $F=0$ via the Lagrange multiplier $B$), while in the premetric formalism, the constitutive law $B=\kappa(F)$ is generally postulated and the excitation is eliminated in favor of the field strength $F$. The difference between~\eqref{eq:constitutive_gauge} and~\eqref{eq:constitutive_gravity} is a fundamental dynamical distinction between gauge theory and gravity, and the main topic of interest here. It allows to go further, and define what is meant by gravity or gauge theory in any manifold setting. As a rephrasing alone, it would be tautologous, but the emphasis is on the structural setup. So, any prototype theory of gravity requires, up to equivalent descriptions,
\begin{enumerate}
	\item The connection $\omega_{ab}$ on the frame bundle $F(M)$, adapted to an inner product $\eta$,
	\item The excitation $B$, reduced to the coframe (hyper)surface basis $*(e^a\wedge e^b)$.
\end{enumerate}
These are essentially requirements that all gravitational quantities are immediately related to spacetime or the underlying manifold. The physically relevant case works around $G=SO(1,3)$, but generally, any other $G$-subbundle defines a gravitational theory with a different sense of relativity. Internal gauge theory, that is electromagnetism and Yang-Mills theory in specific, then introduces connections on a separate principal bundle, and would require
\begin{enumerate}
	\item The connection $A$ on a separate principal bundle $B$,
	\item The excitation $B$, reduced to dual field strength $*F$.
\end{enumerate}
The physically relevant structure group is the Standard Model gauge group, but arbitrary (even non-compact) Lie groups can be considered. Note that the distinction between (internal) gauge theory and gravity has a fundamentally geometric component as the theories apply to different sectors, and is not contingent on the presence of e.g.~first class constraints. The definition of ``gauge theory'' in general can be made precise in various formulations, see e.g.~Trautman~\cite{Trautman:1980,Trautman:1982}, but taken literally, gravity and gauge theory still differ in both the (dynamical) field content and the symmetry content, as discussed in this section. These differences are not as consequential when considering either only in the sense of field theory, but are a subtle obstruction for unification, as it is not clear how (or if at all) the difference or change in geometry should be handled or eliminated.

So far, this has been an overview of standard geometry. These conditions reproduce the standard Lagrangians respectively for gravity or electromagnetism and Yang-Mills theory. In essence, condition 1.~is a requirement on the sector of geometry, and condition 2.~determines the dynamics of the theory, and is a generalized sense of charge and flux conservation per~\eqref{eq:charge_conservation}. If electromagnetism and Yang-Mills theory can be collectively called internal gauge theory, it is tempting to mirror this by calling gravity external gauge theory, specifically when the $B$-field is constrained as described earlier, and motivated by the fact that they are both sprung from the same BF premise, and particularly so as the frame bundle is likewise a principal bundle. However, the more pressing issue is that the $B$ field cannot be constrained uniquely, and could similarly admit a term with the Barbero-Immirzi parameter, or a topological Yang-Mills component. As such, either condition is only justified to the extent that they reproduce standard theory, and can be straightforwardly altered. Generally, the physical relevance of constraining the $B$ field to gravity is well understood, see~\cite{Capovilla:1991kx,Capovilla:2001zi} in particular for early work in this topic, and Plebanski theory~\cite{Plebanski:1977} as a prototypical example. However, the case when the would-be gravitational $B$ field does not inherently take the form of the standard hypersurface basis $e^a\wedge e^b$ or $*(e^a\wedge e^b)$ has not been extensively studied. Of similar importance is the actual geometry where internal and gravitational interactions should situate, and this premise is necessary for even providing a clear interpretation for either sector of interactions. We will follow these questions in section~\ref{sec:unified_structures}.

\subsection{Synchronization}
\label{sec:synchronization}

So far, the unified treatment has been largely trivial, as the frame bundle $F(M)$ of the base manifold and the internal principal bundle $P_\text{int}$ are separate, with a total
\begin{equation}
	P_\text{total}=F(M)\oplus P_\text{int}.
\end{equation}
This is conformant to the Coleman-Mandula theorem~\cite{Coleman_Mandula:1967} which is commonly cited as an obstruction to nontrivial gauge-gravity unification.
However, note that for theories with a larger gauge group, only a vanishing vacuum excitation value of the solder form would signify a unified phase~\cite{Krasnov:2018}. The absence of a nondegenerate metric violates assumptions of the Coleman-Mandula theorem, and the expectation should be that the physics in the unified phase is significantly different from what Poincaré-invariant scattering theory might describe.

Let us briefly add that a nontrivial gauge-gravity unified phase can also manifest as a geometric obstruction. In particular, this appears when trying to construct spacetime directly, as a collection of observers, from defining the set, topology, and smooth structure, and conflating notions of tangent structure and observer states. Each point in the spacetime set $M$ is to be determined by the state of an observer, in some linear space, such that the state vector is also understood as tangent to the underlying manifold, that is both spacetime and the internal fibers. In turn, this requires a choice of topology, which becomes a physical choice how signals are transmitted (synchronized) to the observer. The idea itself of constructing spacetime from some more fundamental notions, ``material points'', has been explored before, cf.~\cite{Finster:2023,Szabados:2025lni}.

However, if the observer state is nontrivial, in that it belongs to a nontrivial representation, no clear distinction is possible between internal, fiberwise directions (corresponding to Yang-Mills theory and electromagnetism), and external, spacetime directions (corresponding to gravity). If the state is nontrivial, it is simply not possible to define the underlying spacetime 4-manifold. However, note that this conclusion is contingent on a change in interpretation. It is similarly reasonable to consider gravity as a (metric, connection, or otherwise) field on a background 4-dimensional spacetime, and not as directly identified with the geometry of spacetime. More importantly, this interpretation requires a change in how observers are defined. Rather than modelled as unit vector fields on the base manifold, they are now promoted to points defining the spacetime, with an attached (linear) structure. Similarly, there is a variety of options when identifying (parts of) the observer state with gauge or other fields. Most obviously, it is possible to use the gauge field at a point as a single vector, which would require introducing gauge transformations to define equivalent states.

The standard description of relativistic observers considers the collection of future timelike vector fields, or is provided by Cartan geometry in terms of model spaces~\cite{Gielen:2013, Hohmann:2016}. Forgoing the background spacetime geometry and taking the point-like observer as a primitive notion, the observer would be naturally modeled as a triplet
\begin{equation}
	o=(\tau,v,\rho),
\end{equation}
with proper time $\tau\in\mathbb{R}$, and the (internal or external) observer state $v\in V$ as a vector in the representation $\rho$ of some symmetry group $G$. In principle, $\tau$ can be included in $v$ by appending the additive group $(\mathbb{R},+)$ to $G$. So far, $o$ can be understood as defining a point $p$ in the total bundle tangent space, $v\in T_p P$. The collection of all observers $o_i$ defines the spacetime set $M$. Proper definition requires more careful handling, in that the observers should non-colliding, and it is assumed that only one observer can be at any given spacetime point simultaneously, but this will not be necessary for the argument of discerning gauge-gravity geometry.

As discussed, for standard geometry the spacetime set $M$ should have more structure so that it is a 4-dimensional pseudo-Riemannian manifold. In particular, $M$ requires a topology, which determines part of the model how signals are transmitted. In this sense, signals can be interpreted as similar triplets transmitted between observers, by some dynamical principles. The physical requirement is identifying signal transmission with continuity, which introduces a topology onto $M$: Zeeman-Göbel or Alexandrov topologies would be more immediately causal, while the manifold topology requires a metric in addition for causal structure. Let us assume that $M$ is a 4-manifold, which is a particularly convenient setting for standard physics. The aim is to describe spacetime as a field of observers, so each $p\in M$ should simultaneously correspond to some $p\equiv o$. The manifold structure of $M$ introduces its own tangent linear structure $T_p M$ at each point. The argument is that $T_p M$ should be embedded into the observer state space $V$, conflating the observer state structure and tangent structure. As an almost trivial contradiction, if the unified phase is nontrivial,
\begin{equation}
	V\neq V_\text{ext}\oplus V_\text{int},
\end{equation}
then clearly an identification $V_\text{ext}\equiv T_p M$ is impossible as it is simply assumed to not exist, and this can be argued to constitute a contradiction. The resolution in this construction is that the 4-dimensional manifold structure should not be forced upon the unified phase. In standard spacetime terms, it is sufficient to consider only the trivially broken phase $F(M)\oplus P_\text{int}$. However, note that this is a highly model-dependent issue, and this class of examples only works when conflating underlying geometry and observer state notions, resulting in an almost-trivial contradiction. Rather, the unified phase has to be studied generally on a case-by-case basis.

The observer line of thought can be pushed further, reinterpreting and redefining common quantities and operations in a different setting. The identification with continuity (thus topology) and signal-transmission was already mentioned. To compare signals and observer states, synchronization would be a choice of $G$-action
\begin{equation}
	v'=g\cdot v,
\end{equation}
including the linear difference $v'=v+v_\Delta$. The space of synchronizable (physical) signals with respect to the $G$-action would be the orbit $G\cdot v$, while $V$ would be the space of all possible incoming information. The space of actual signals
\begin{equation}
	O_\text{In}(\tau)=\{o_i\vert i\in\mathcal{I}\}
\end{equation}
would depend on the dynamical model, and is the only information available to reconstruct external geometry, suggesting an information-theoretic construction of interactions. A continuous model of signal transmission would introduce a sequence of distributions on the state space. However, these developments will currently be deferred, with the focus instead on structural differences between gravitational and internal gauge interactions.

\section{Unified structures}
\label{sec:unified_structures}

\subsection{BF theory}
\label{sec:BF_theory}

Let us return to the manifold setting and assume the standard setup for field theory, namely that there is an underlying spacetime 4-manifold $M$ with a principal $G$-bundle, which should include both gravity and all gauge interactions. In the broken phase, gravity should be described by the frame bundle of the base manifold, while internal gauge theory is in a separate principal bundle. So,
\begin{equation}\label{eq:BF_structure_general}
	G=SO(1,3)\times G_\text{int},
\end{equation}
where the $G_\text{int}$ sector should correspond to the Standard Model or a GUT group. Then, the generic covariant candidate for field theory dynamics begins with the BF theory action
\begin{equation}\label{eq:BF_action}
	S_\text{BF}=\int_M \Tr_G (B\wedge F).
\end{equation}
Some comments are in order. Here, $A$ is a $G$-connection with curvature $F[A]$, and $B$ is a $\mathfrak{g}$-valued 2-form. As both $B$ and $F$ transform in the adjoint, $F\to g F g^{-1}$ and $B\to g B g^{-1}$, the action is gauge invariant. More general actions can be considered, such as when $B$ is in a different Lie algebra $\mathfrak{h}$ and mapped to $\mathfrak{g}$ through some homomorphism, or $\Tr_G$ is instead some other invariant polynomial. However, we will restrict to the simplest case~\eqref{eq:BF_action}. Physically, it should be emphasized that BF theory is simply elevating the gauge excitation $B\equiv H$ (or, the dual field strength) to be an independent field, the closest option to a Palatini formalism for generic gauge theory. Without any additional potential or coupling it is a purely topological theory, without bulk degrees of freedom~\cite{Cattaneo_et_al:1995,Evnin_et_al:2024}.

So far, there is no immediate relation to gravity, to the base spacetime manifold, or to Yang-Mills theory, but bar constraints, it is close to the minimum to write down any prototype for the dynamics of the connection. Constructively, Lagrangian dynamics is sufficiently generic, as the action simply weighs trajectories. For dynamical connections $A$, the gauge-invariant Lagrangian must include a term with the field strength $F$, and for a top rank Lagrangian form, the remainder is the generic complement $B$, the excitation, which is to be constrained. Consider table~\ref{table:excitations} for the lowest order options. Without introducing higher-order terms or additional fields, the choices capable of relativistic dynamics are unique.

\begin{table}[h]
	\caption{Selection of lowest order gauge-covariant constitutive laws for the excitation $B$ available in 4 dimensions, without introducing additional field content; see also~\cite{Rezende:2009sv,Hughes:2024}. All topological terms would modify the boundary and canonical structure, with possible quantum effects; see also~\cite{Bott:1982xhp}. This selection is limited by the basic fields and geometric operations $\wedge$ and $*$. A more generic $B\sim\kappa(F)$ would provide principal, skewon and axion-like contributions~\cite{Hehl_Obukhov:2003,Hehl:2004zz}, while the addition of $\Der e^a\wedge\Der e_a$ would allow for the Nieh-Yan term. Similarly, other dimensions offer different variety, e.g.~(standard) Chern-Simons theory in three dimension, while $B=F$ is no longer of maximal rank and cannot produce a Lagrangian $n$-form.}\label{table:excitations}
	\begin{tabular}{ c c c }
		\toprule
		$B=$ & Contribution & Interpretation \\ \midrule
		$B$ & Pure BF theory & Topological, unbroken premetric phase \\
		$*(e^a\wedge e^b)$ & Gravity & Direct GR-like contribution \\
		$e^a\wedge e^b$ & Immirzi parameter, Holst term & Various: quantum area measure, black hole entropy, etc. \\
		$*R_{ab}[\omega]$ & Euler class, Kretschmann scalar &
		Topological invariant \\
		$R_{ab}[\omega]$ & Pontryagin class & Topological invariant \\
		$*F[A_\text{int}]$ & Maxwell/Yang-Mills theory & Standard gauge theory \\
		$F[A_\text{int}]$ & Topological Yang-Mills theory & Topological invariant \\
		\bottomrule
	\end{tabular}
\end{table}

With a quadratic potential (or, the cosmological constant term)
\begin{equation}\label{eq:1stYM}
	S_\text{1\textsuperscript{st} YM}=\int_M\Tr_G \bigg(B\wedge F + \frac{g^2}{2}B\wedge *B\bigg)
\end{equation}
it becomes the first order formulation of Yang-Mills theory, which is both classically and quantum-mechanically fully equivalent to the standard, second order formulation
\begin{equation}
	S_\text{gauge} = \int\frac{1}{2g^2}\Tr_G(F\wedge *F).
\end{equation}
More generic potentials are possible, and have been studied as well~\cite{Krasnov:2009,Celada:2016}.

Currently,~\eqref{eq:BF_action} and~\eqref{eq:1stYM} give the premise for a gauge-gravity unified theory with a trivially organized symmetry group~\eqref{eq:BF_structure_general}. All subconnections and subexcitations are collected into the total fields
\begin{align}
	B&=B_{ab}T^{ab}_\text{ext} + B^I T^I_\text{int},\\
	A&=\omega_\text{ext} + A_\text{int}.
\end{align}
Similarly, because of~\eqref{eq:BF_structure_general}, the external and internal generators commute,
\begin{equation}
	[T^{ab}_\text{ext}, T^I_\text{int}]=0,
\end{equation}
so
\begin{equation}
	\Tr(B_\text{ext}\wedge F[A_\text{int}])=\Tr(F[A_\text{int}]\wedge B_\text{ext}).
\end{equation}
In particular, this also implies that the field strengths split into
\begin{equation}
	F[A]=R[\omega_\text{ext}] + F[A_\text{int}].
\end{equation}
Most importantly, because the generators of any semi-simple Lie algebra must be traceless, the action reorganizes to
\begin{equation}
	S=\int\frac{1}{2g^2}\Tr_{SO}(B_\text{ext}\wedge R[\omega_\text{ext}]) + \frac{1}{2g^2}\Tr_{G_\text{int}}(B_\text{int}\wedge F[A_\text{int}]),
\end{equation}
with gravity and internal gauge theory separated. Note the quadratic potential $B\wedge*B$ decomposes analogously. If the internal gauge group is similarly organized as
\begin{equation}\label{eq:internal_symmetry}
	G_\text{int} = G_1\times G_2\times\ldots\times G_n,
\end{equation}
so that for $i\neq j$
\begin{equation}
	[T^I_{G_i},T^J_{G_j}]=0
\end{equation}
and
\begin{equation}\label{eq:product_trace}
	\Tr(T^I_{G_i} T^J_{G_j})=0,
\end{equation}
then the internal contribution also sums into
\begin{equation}
	\Tr(B_\text{int}\wedge F[A_\text{int}])=\Tr\bigg[\sum_i B_{G_i}\wedge F[A_{G_i}]\bigg].
\end{equation}
In particular, this permits a single Abelian component, as in the Standard Model
\begin{equation}
	U(1)_Y\times SU(2)_L\times SU(3).
\end{equation}
Finding the most suitable structure (group, algebra, etc.) is one of the central questions of unification, but here the primary interest is in the distinction between the gravitational excitation $B_\text{ext}$ and the internal excitation $B_\text{int}$.

Gravity requires some mechanism to restrict the excitation to the hypersurface basis $*(e^a\wedge e^b)$, while the internal excitation should map to the dual field strengths $*F[A_i]$ weighed by coupling constants $g_i$. In the action~\eqref{eq:BF_action}, there must be a process to map
\begin{subequations}\label{eq:constr_process}
	\begin{align}
		B\big\vert_{G_i}&=\frac{1}{2 g_i^2}*F[A_i],\\
		B\big\vert_\text{ext}&=\frac{1}{2\kappa}*(e^a\wedge e^b),
	\end{align}
\end{subequations}
where $G_i$ would refer to internal symmetry groups in~\eqref{eq:internal_symmetry}. In premetric terms,~\eqref{eq:constr_process} are the constitutive laws of the theories. As argued earlier, the presence of the hypersurface basis $*(e^a\wedge e^b)$ vs.~the field strength $*F$ is also the lowest order distinction in $F$ and $e^a\sim g_{\mu\nu}$ between an external and an internal gauge theory, that is between gravity and Yang-Mills theory or electromagnetism. At the same time, note the placement of the Hodge dual in~\eqref{eq:constr_process} is not strictly necessary, when considering instead $B'\equiv*B$ as the fundamental field, as in
\begin{equation}
	S_{B'F}=\int \frac{1}{2g^2}B'\wedge *F.
\end{equation}
This sacrifices the topological nature of the theory by introducing causality, the metric and gravitational structure earlier and moving the Hodge star out of the constraints, but the basic internal-external distinction of $e^a\wedge e^b$ vs.~$F$ remains unchanged.

For simplicity, let us consider a generic (compact) $G_\text{int}$ with only one constant $g$, that is
\begin{equation}
	B\big\vert_{G_\text{int}}=\frac{1}{2 g^2}*F[A_\text{int}],
\end{equation}
which could be further broken by some additional scheme. There is not an unlimited amount of options how to introduce the constraints~\eqref{eq:constr_process}. Generally, these relations could be simply structural, that is postulated or substituted for through some identities, or derived as a component of some larger set of equations of motion (that is, a nontrivial part of the equations of motion of a larger object), or appended separately by Lagrange multipliers (essentially, a trivially separated part of some set of equations). There is variety in the exact form of constraints, but most directly, everything can be collected into a total Lagrange multiplier term
\begin{equation}
	\int\Tr_G(\lambda\wedge(B-C[\phi])).
\end{equation}
Because of this, $\lambda$ must be adjoint-valued as well, otherwise the trace vanishes. Most obviously, a direct restriction
\begin{subequations}
	\begin{align}
		C\vert_{G_\text{int}}&=\frac{1}{2g^2}*F[A_\text{int}],\\
		C\vert_{SO}&=\frac{1}{2\kappa}*(e^a\wedge e^b),
	\end{align}
\end{subequations}
so
\begin{equation}
	C[A,e^a]=*F[A_{G_\text{int}}] + *(e^a\wedge e^b)\in \Omega^2(\mathfrak{g}_\text{int}) \oplus \Omega^2(\mathfrak{so}(1,3)).
\end{equation}
The quadratic potential would trivialize the internal gauge theory constraint, as it becomes unnecessary. That is, only the gravity constraint is required,
\begin{equation}
	\Tr_{SO}(\lambda\wedge(B - C[e^a]))=\lambda_{ab}\wedge(B^{ab} - *(e^a\wedge e^b)).
\end{equation}
The $B^{ab}\wedge*B_{ab}$ term would become the proper cosmological constant, with important implications for quantum gravity, as the construction of the Kodama or Chern-Simons state~\cite{Kodama:1990,Magueijo:2021,Alexander:2024} requires its presence. Constraints for gravity in BF terms have been quite extensively studied in literature, early comprehensive work, besides Plebanski, include Capovilla \textit{et al.}~\cite{Capovilla:1991kx,Capovilla:2001zi}, which also includes the addition of e.g.~the Immirzi parameter. However, the simultaneous (unified) treatment has been less studied, even the description of basic structure; cf.~the review~\cite{Krasnov:2018}. To reiterate, several effects appear simultaneously: in the unified phase, dynamics are described by the generic topological BF theory. To get realistic physics, the $B$-field is broken to different values for gravity and for internal gauge theory, respectively the Einstein-Cartan/metric-affine hypersurface element or the Yang-Mills/electromagnetic dual field strength. Furthermore, gravitational fields are to situate in the geometry of the underlying manifold, respectively the coframe structure (reduced to, expectedly, $SO(1,3)$) and as a Lorentz connection on the frame bundle, while internal gauge theory is on a separate principal bundle. In future work, it would be interesting to realize this constraining process in some dynamical terms, and not as just Lagrangian constraints, although it would be unlikely to derive particularly interesting phenomenological signatures. As a particular line of thought, it would appear feasible to introduce spacetime as a hypersurface in a higher-dimensional space cf.~e.g.~\cite{PonceDeLeon:2001un}, so the broken phase would be attached boundary-wise. But currently, the focus is only in the broken phase.

So, in symmetry breaking terms, these constraints break a transformation between the different gauge sectors and instead impose a polynomial symmetry in $e^a$. Because of this, the initial BF theory~\eqref{eq:BF_action}, written as
\begin{equation}
	S_\text{BF}=\int_M \frac{1}{2g^2}\Tr_G((B_\text{int}+B_\text{ext})\wedge(F[A_\text{int}] + R[\omega_\text{ext}])),
\end{equation}
is invariant under the (vectorized) transformation
\begin{equation}\label{eq:B_rotation}
	\begin{gathered}
		\begin{pmatrix}
			B_\text{int}\\ B_\text{ext}
		\end{pmatrix}
		\to
		\begin{pmatrix}
			a_1&a_2\\
			a_3&a_4
		\end{pmatrix}
		\begin{pmatrix}
			B_\text{int}\\ B_\text{ext}
		\end{pmatrix}\\
		a_1 + a_3 = 1,\ a_2 + a_4 = 1,
	\end{gathered}
\end{equation}
so
\begin{equation}
	S_\text{BF}\to\int_M \frac{1}{2g^2}\Tr_G\Big(((a_1 + a_3)B_\text{int}+ (a_2 + a_4)B_\text{ext})\wedge(F[A_\text{int}] + R[\omega_\text{ext}])\Big)=S_\text{BF}.
\end{equation}
This is not a proper rotation, in the sense of the excitations $(B_\text{int},B_\text{ext})$ forming a doublet of the rotation group $SO(2)$, and it is not part of a Lie group, as the transformation matrix can be singular. Equation~\eqref{eq:B_rotation} can also be understood as an invariance transformation of the trace, e.g.~a similarity transformation of the matrix trace
\begin{equation}
	B_\text{int} + B_\text{ext} = \Tr
	\begin{pmatrix}
		B_\text{int}&0\\
		0&B_\text{ext}
	\end{pmatrix}=
	\Tr\bigg[P^{-1}
	\begin{pmatrix}
		B_\text{int}&0\\
		0&B_\text{ext}
	\end{pmatrix}
	P\bigg],
\end{equation}
where $P$ is an arbitrary $GL(2)$ matrix. The field strengths can only be put into a formal doublet, as generally it is not possible to scale $F[A]\to a F[A]$ by a transformation of the basic variable $A$. That is, it would be a transformation of the full space of 2-forms $\Omega^2(\mathfrak{g}_\text{int})\oplus\Omega^2(\mathfrak{so}(1,3))$, which would include the curvatures.

It is possible to take an effective field theory point of view, and instead consider all possible Lagrangian terms up to some order~\cite{Rezende:2009sv}. In this case there is no symmetry breaking, except for the values of the arbitrary constants that could be added. The excitation would be the total (direct) sum
\begin{equation}
	B = \alpha_1*(e^a\wedge e^b) + \alpha_2 (e^a\wedge e^b) + \alpha_3 *R^{ab}[\omega_\text{ext}] + \alpha_4 R^{ab}[\omega_\text{ext}] 	+ \alpha_5 *F[A_\text{int}] + \alpha_6 F[A_\text{int}],
\end{equation}
and the terms trivially separate per the traceless generator condition discussed earlier. That is, for example
\begin{equation}
	\Tr_G(*(e^a\wedge e^b)\wedge F[A_\text{int}])=*(e^a\wedge e_a)\wedge F^I[A_\text{int}]\Tr_G T^I = 0.
\end{equation}
We return to the issue that unification is highly model-dependent, as either interpretation has separate merit.

\subsection{Equivalence and duality}
\label{sec:duality}

Although the initial premise in the field content in the symmetry breaking or constraint point of view is different compared to simply postulating the final actions, the resolution to the constraint surface is not physically different. The classical theory passes through the equations of motion, while the quantum theory would directly integrate the Lagrange multiplier part of the path integral. To be precise, the gauge theory path integral is incomplete without removing gauge freedom, while the Lorentz group is non-compact and gravity is not renormalizable leading to a separate variety of problems. Nevertheless, this does not change the premise how quantum theory would behave, where
\begin{equation}
	\int\mathcal{D}\lambda\exp\bigg[i\int\lambda\wedge C(A,B)\bigg]=\delta(C(A,B)),
\end{equation}
where the zeros
\begin{equation}
	C(A,B)=0\Rightarrow B=g(A)
\end{equation}
would provide
\begin{equation}\label{eq:basic_equivalence}
	\begin{aligned}
		Z[A,B,\lambda]&=\int\mathcal{D}A\ \mathcal{D}B\ \mathcal{D}\lambda\exp\bigg[i\int B\wedge F + \lambda\wedge C(A,B)\bigg]\\
		&=Z[A]=\int\mathcal{D}A\exp\bigg[i\int g(A)\wedge F\bigg]
	\end{aligned}
\end{equation}
for the internal gauge part of the BF path integral. Note that the $B$ measure was eliminated with the Dirac delta; for many relevant $C(A,B)$ the solution is unique, e.g.~${\frac{1}{2g^2}*F[A]}-B=0\Rightarrow B={\frac{1}{2g^2}*F[A]}$. Otherwise, the solution requires introducing a sum over the zeros,
\begin{equation}
	\delta(C(A,B))=\sum_i\frac{\delta(B-A_i)}{C'(A,B)},
\end{equation}
and would also appear in~\eqref{eq:basic_equivalence}. Gauge freedom can be removed with e.g.~the BRST formalism, which would add (anti)field content, separate from the rest of the action. This is roughly how it is possible to prove quantum equivalence between the first and second order Maxwell theory~\cite{Lavrov:2021}, or how to derive duality relations in Maxwell-Chern-Simons~\cite{Armoni:2022xhy}. For any substantial difference, there must be further modification, in either the field content, the constraint structure (e.g.~permitting a larger class of solutions), or the constraining process itself, and indeed, the issue is that $B$ itself can be interpreted as a Lagrange multiplier.

Consider a significantly weaker differential condition on the excitation. The Bianchi identity is a necessary condition for a 2-form to be the curvature, but it is not sufficient. So, the prototype Bianchi identity or the homogeneous set of Maxwell or Yang-Mills equations
\begin{equation}
	\Der*B=0,
\end{equation}
has $B=\frac{1}{2g^2}*F$ (or any arbitrary other curvature $F'\neq F$) as a particular solution, but permits the entire class of covariantly constant 2-forms. Note the coupling constant is introduced just as an arbitrary constant during integration. The issue is in the back-reaction $B\leftrightarrow F$ of both dynamical variables. This constraint is easily introduced by the addition of an adjoint-valued Lagrange multiplier
\begin{equation}
	S_{\Der*B}=\int \Tr(B\wedge F + \lambda_{\Der*B}\wedge\Der*B),
\end{equation}
with the equations of motion
\begin{subequations}
	\begin{align}
		\delta_{\lambda_{\Der*B}}&:\Der*B=0,\\
		\delta_B&:F+*\Der\lambda_{\Der*B}=0\\
		\delta_A&:\Der B +[\lambda_{\Der*B},*B]=(J).
	\end{align}
\end{subequations}
The first equation is simply the prototype Bianchi identity, but the second spoils the system. In the case of $U(1)$, the primary field strength $F$ simply obtains a vacuum duality relation $*F=\der \lambda_{\Der*B}\Rightarrow \der*F=0$, so a $J\neq0$ case does not permit a $B\sim*F$ solution. The essence of the issue is that $B$ can also be interpreted as a Lagrange multiplier, here identifying two different curvatures,
\begin{equation}
	S_{\Der*B}=\int\Tr(B\wedge(F+*\Der\lambda))
\end{equation}
In non-Abelian theory, $[\lambda_{\Der*B},*B]=\lambda_{\Der*B}\wedge*B - *B\wedge\lambda_{\Der*B}\neq0$ persists as a background structure, while $F=-{*\Der\lambda_{\Der*B}}$ becomes a constraint on the field strength as it is not generically true. Therefore, this type of weaker constitutive law can be dismissed as physically unviable.

As another nontrivial option it is tempting to attempt what could be called a ``spontaneous'' breaking to the physical constitutive law. This must necessarily extend from the standard understanding, where spontaneous symmetry breaking only applies to global Lie group (and suitably generalized, e.g.~higher form) symmetries, while the constitutive law is a functional identification. So, the aim is to devise a potential with a nontrivial structure of the minima, where the physical constitutive law $B=\kappa(A)$ would be one possible solution to the following $B$-equations of motion, in addition to other, presumably nonphysical solutions. As the $B$-field is to be auxiliary, it should be eliminated from the list of dynamical variables: in~\eqref{eq:1stYM}, this can be done immediately through Gaussian integration. However, for a set of multiple zeros, another choice (``breaking'') is required, similar to a ``spontaneous'' process. However, it should be emphasized that this is also the extent of similarity to spontaneous symmetry breaking in the usual sense, as it is not any obvious Lie group symmetry that is broken. Altogether classically, mathematically we will try to construct $B$-potentials with multiple nontrivial solutions, but without introducing additional (Lagrange multiplier) fields, while physically, this choice corresponds to choosing the constitutive law and actually defining the dynamics of the gauge theory, e.g.~in electromagnetic terms fixing details of charge conservation, light cone structure and wave propagation~\cite{Hehl_Obukhov:2003}. But more specifically, as will be illustrated, the issue is not that it is difficult to construct potentials where the constitutive law between $B$ and $F$ is degenerate or not unique (so a particular choice can be called ``spontaneous''), but rather that any obvious candidate will introduce back-reaction from field strength terms, which cannot be physically taken as a background. The issue is that the constitutive law minimum is dynamical, in the sense that it should relate two dynamical quantities $B$ and $F$, while interpreting $B$ as the gauge excitation is only sensible when it is the dual to the field strength $F$, thus fixing the $B\wedge F$ term.

It is not difficult to position $B-*F$ at the minimum of some polynomial potential when forgoing BF theory. Consider the $\mathfrak{g}$-valued 2-form $b$. As the basic field is an antisymmetric tensor $b_{ab}$ in 4 dimensions, there is greater variety in contracting a quartic term, cf.~all the contractions in $b_{i_1}{}^{j_1} b_{i_2}{}^{j_2} b_{i_3}{}^{j_3} b_{i_4}{}^{j_4}$ and all combinations of $*$, $\lrcorner$ and $\wedge$. But the most obvious quartic potential is
\begin{equation}
	V_b=\Tr(\mu_1 b\wedge*b + \mu_2 (b\wedge *b)^2 )
\end{equation}
where
\begin{equation}
	(b\wedge *b)^2=(b\wedge *b)(*(b\wedge *b)),
\end{equation}
neglecting other combinations like $(b\wedge b)(*(b\wedge b))$. The potential extremizes at
\begin{equation}\label{eq:quartic_constitutive}
	\delta_b V_b =0\Rightarrow *b (\mu_1 + 2\mu_2*(b\wedge *b))=0\Rightarrow
	\begin{cases}
		b=0,\\
		b\wedge *b=-*\frac{\mu_1}{2\mu_2}.
	\end{cases}
\end{equation}
The specific behaviour depends on the particular choice of $\mu_1$ and $\mu_2$, while $B$ and $F$ can be inserted through Lagrange multipliers. Unfortunately, neither case of~\eqref{eq:quartic_constitutive} is satisfactory. Positioning the constitutive law at either minimum would respectively require the constraints
\begin{subequations}
	\begin{align}
		b&=B-\frac{1}{2g^2}*F,\ \text{or}\\
		b\wedge *b&=B\wedge*B,\ *\frac{\mu_1}{2\mu_2}=\frac{1}{2g^2}F\wedge*F,
	\end{align}
\end{subequations}
and, when substituting these identities back to the potential, both would lead to a nonlinear gauge theory.

So, let us consider the generic case for BF internal gauge theory. For the premetric interpretation to work, the excitation $B$ must appear as a complement to field strength, and $F$ in turn should not appear in the potential; otherwise, there will be back-reaction to the inhomogeneous equation, leading to either inconsistency with observation or fine tuning. The generic action
\begin{equation}
	S_{BF+V}=\int \Tr_G(B\wedge F + V(B) - A\wedge J(\phi))
\end{equation}
has the equations of motion
\begin{subequations}
	\begin{align}
		\Der B &= J,\label{eq:YM_proto}\\
		F &= -\frac{\delta V}{\delta B}.\label{eq:constitutive_proto}
	\end{align}
\end{subequations}
The prototype Yang-Mills equation~\eqref{eq:YM_proto} must be physical as it couples to particle currents, while the constitutive equation~\eqref{eq:constitutive_proto} fixes the constitutive law $B=\kappa(F)$ of the gauge theory, and must be resolved as auxiliary to eliminate the excitation. If the potential $V$ only includes the excitation $B$, then clearly the field strength $F$ is uniquely identified with some function of $B$. However, the inversion
\begin{equation}
	F=f(B)\Rightarrow B=f^{-1}(F)
\end{equation}
need not be unique. This is one aspect of spontaneous symmetry breaking which allows it to be generalized to functional form. The aim is to reconstruct a potential $V(B)$ such that equation~\eqref{eq:constitutive_proto} has multiple solutions,
\begin{equation}
	B=
	\begin{cases}
		\frac{1}{g^2}*F,\\
		\quad \vdots
	\end{cases}
\end{equation}
Generally, this would be a Lagrangian reconstruction (``integration'') problem\footnote{In related terms, this is a part of the inverse problem of variational calculus, and could be studied in terms of the Vainberg-Tonti Lagrangian and variational completion~\cite{Voicu_Krupka:2015}.}, reconstructing the initial potential from its variational derivative w.r.t~$B$ and given solution points. However, it is simple to show that this cannot appear from a polynomial potential. In component form, let $*F\sim\tilde{F}_i{}^j$, so for a polynomial constitutive law
\begin{equation}
		\tilde{F}_i{}^j=a_0 C_i{}^j + a_1 B_i{}^j + \ldots + a_n (B_i{}^j)^n,
\end{equation}
where $C_i{}^j$ is a constant background field. The physical constitutive law $B_i{}^j=\frac{1}{2g^2}\tilde{F}_i{}^j$ should be a particular solution, so substituting back into the equation results in
\begin{equation}
	\tilde{F}_i{}^j=a_0 C_i{}^j + a_1 \frac{1}{2g^2}\tilde{F}_i{}^j + \ldots + a_n \bigg(\frac{1}{2g^2}\tilde{F}_i{}^j\bigg)^n.
\end{equation}
This is generally inconsistent, as it is interpretable as a constraint for $\tilde{F}_i{}^j$. That is, we are provided an additional identity. The quadratic potential case only retains linear terms, so it trivially annihilates, but otherwise there is not an \textit{a priori} reason why powers of $\tilde{F}_i{}^j$ of different degree should eliminate each other. To get a slightly better view of the solutions, consider that for polynomial rings, so real and complex matrices $T_i{}^j$ in particular, the matrix polynomial factorizes as
\begin{equation}
	a_0 + a_1 T + \ldots + a_n T^n = c(T- r_1\mathds{1})\ldots(T- r_m\mathds{1}).
\end{equation}
Applied to the constitutive law,
\begin{equation}
	(\tilde{F}_i{}^{a_1} - r_1\delta_i^{a_1})\ldots(\tilde{F}_{a_m}{}^j - r_m\delta_{a_m}^j)=0.
\end{equation}
Non-polynomial interactions would generally not be renormalizable, so a spontaneous constitutive law would not be physically viable. This is sufficiently generic as well, because in a power law expansion the polynomial potential would be a truncation of the series.

For comparison, consider the fully generic case as field shifted to a deviation from the constitutive law and define the potential as a deviation from the solvable quadratic potential,
\begin{subequations}
	\begin{align}
	B&\to*F/g^2 + b,\\
	V(B)&\to\frac{g^2}{2}B\wedge*B + V_\delta(B)
	\end{align}
\end{subequations}
so the action reads
\begin{equation}
	S_{BF+V}\to\int\Tr_G\bigg(\frac{1}{2g^2}F\wedge*F  
	+ \frac{g^2}{2}b\wedge*b
	+ V_\delta(*F/g^2 + b) - A\wedge J(\phi)\bigg).
\end{equation}
This yields
\begin{subequations}
	\begin{gather}
		\frac{1}{g^2}\Der*F + \frac{\delta V_\delta(*F/g^2 + b)}{\delta A} = J(\phi),\\
		g^2*b + \frac{\delta V_\delta(*F/g^2 + b)}{\delta b}=0.\label{eq:deviation_constitutive}
	\end{gather}
\end{subequations}
In order not to have any contribution to the Yang-Mills equations, we get the additional constraint
\begin{equation}
	\frac{\delta V_\delta(*F/g^2 + b)}{\delta A}=0,
\end{equation}
while~\eqref{eq:deviation_constitutive} should not introduce any additional conditions on $F$. The constraint~\eqref{eq:deviation_constitutive}, however, is harsh, as it requires the deviation from the quadratic potential, evaluated at $*F/g^2 + b$, to not have any dependence on $A$. Generally, this is not viable, outside of the trivial $V_\delta=0$. 

\subsection{Nontrivial constitutive laws}
\label{sec:constitutive_laws}

The final structural question to consider is the relation between the gravitational excitation $B_\text{ext}$ and the internal gauge excitation $B_\text{int}$. In principle, as the interactions are independent, there is no \textit{a priori} expectation of a relation between the values either is fixed to. On physical grounds and as argued earlier, the possibilities are not endless either, and there are definite and unique preferences to the Lagrangian structure. Nevertheless, as there has been increasing interest in theories of modified gravity, it is natural to ask whether a theory of gravity with a modified gravitational excitation would induce a modification in the internal gauge excitation. Although the answer is of course negative, it does permit to introduce the mappings between the spaces of $\mathfrak{g}_\text{ext}$ and $\mathfrak{g}_\text{int}$-valued $n-2$ forms as a separate concept in its own right. As these are mappings between constitutive laws and excitations, they will be called constitutive mappings (or morphisms, when suitable).

As a specific example, there was a recent ``Khronon'' Lorentz gauge theory of gravity~\cite{Zlosnik:2018qvg}, which introduced the coframe $e^a\equiv\Der\tau^a$ through a Lorentz vector field $\tau^a$ and the self-dual spin connection $\omega_{ab}$; see also~\cite{Koivisto:2025} for further 3+1 and black hole specifics. Then, the action
\begin{equation}\label{eq:khronon}
	S_{\tau^a}=\frac{i}{2\kappa}\int\Der\tau^a\wedge\Der\tau^b\wedge R^+_{ab}
\end{equation}
reproduces General Relativity with cosmological dark matter dust when spontaneously breaking Lorentz symmetry with $\tau^2<0$. This is an example of a nontrivial gravitational constitutive law $*(\Der\tau^a\wedge\Der\tau^b)$, instead of the usual hypersurface basis element $*(e^a\wedge e^b)$. Note that the Hodge dual $*$ and the Lorentz group dualization $\star$ coincide on the surface basis, and can be absorbed into the self-dual curvature~\cite{Giulini:1994}. Furthermore, as the ``Khronon'' $\tau^a$ mimics the contact vector in (de Sitter) Cartan geometry~\cite{Westman:2012zk,Westman:2013mf,Westman:2014yca} and is directly interpretable as describing the observer's state, it is tempting to attempt a unification procedure with special consideration of the particular vector (or, Lorentz-valued scalar) field $\tau^a$, coframe $e^a=\Der\tau^a$, and excitation $B_\text{ext}=*(\Der\tau^a\wedge\Der\tau^b)$.

Let us give a different perspective why any such approach is unsuccessful in the standard formulation of gauge theory and gravity, and would instead either require a different geometric premise (as in Cartan geometry, or the observer state structure discussed earlier), or a postulated nontrivial relation between the excitations (which would be more \textit{ad hoc} than hoped). The issue is that there is not a canonical mapping $B_\text{ext}\leftrightarrow B_\text{int}$. This can be expressed in several ways. Most directly, this can be expressed by showing that the following commutative (constitutive) diagram is not canonically spanned or well-defined:
\begin{equation}\label{eq:constitutive_diagram}
	\begin{tikzcd}[column sep=tiny]
			&	\underset{B}{\Omega^{n-2}(\mathfrak{g})}\arrow[dl, "C_\text{int}"']\arrow[dr, "C_\text{ext}"]	&	 \\
		\underset{B_\text{int}=*F}{\Omega^{n-2}(\mathfrak{g}_\text{int})}\arrow[rr, "\phi_\text{i-e}", shift left=0.5ex] 	&		&	\underset{B_\text{ext}=*(e^a\wedge e^b)}{\Omega^{n-2}(\mathfrak{g_\text{ext}})}\arrow[ll, "\phi_\text{e-i}", shift left=0.5ex]
	\end{tikzcd}
\end{equation}
The mappings are between bundles of differential forms, but they are underdefined, as only the restrictions to the sections $B$, $*F$ and $*(e^a\wedge e^b)$ are determined. For example, $C_\text{ext}$ is arbitrary as long as
\begin{equation}
	C_\text{ext}(B_\text{full})=*(e^a\wedge e^b).
\end{equation}
In the broken phase, where gravity and internal gauge theory are separated, the gauge excitation of the theory should be
\begin{equation}
	\begin{aligned}
		B=B_\text{int} + B_\text{ext} &=\frac{1}{2g^2}*F + \frac{1}{2\kappa}*(e^a\wedge e^b)\\
		&= C_\text{int}(B) + C_\text{ext}(B)\\
		&=\phi_{e-i}(*(e^a\wedge e^b)) + \frac{1}{2\kappa}*(e^a\wedge e^b)\\
		&=\frac{1}{2g^2}*F + \phi_{i-e}(*F),
	\end{aligned}
\end{equation}
by hand or through some suitable mechanism. But off-shell, the sections are arbitrary, and $C_\text{ext}, C_\text{int}, \phi_\text{i-e}, \phi_\text{e-i}$ hold any meaning only as mappings between functional expressions (that is, between function spaces, e.g.~polynomials, in a meta-characteristic sense). Otherwise, the off-shell sections are entirely arbitrary. This is another phrasing of the constitutive issue: the mappings are only between three individual, off-shell arbitrary sections, and these mappings cannot be meaningfully extended to other sections. Alternatively, because the fields remain $n-2$ forms, it is also feasible to think of these relations as mappings between different Lie algebras.

The bundles $\Omega^{n-2}$ in the constitutive diagram could be replaced by the space of solutions $\mathcal{P}\subset\Gamma(\Omega^{n-2}(\mathfrak{g}_\text{ext})\oplus\Omega^{n-2}(\mathfrak{g}_\text{int})\oplus\ldots)$, that is the covariant phase space~\cite{Harlow:2019yfa}, as this would allow to map solutions to solutions $B_\text{ext}\leftrightarrow B_\text{int}$. This would not resolve the issues off-shell, while a modified constitutive law might simply change the initial value problem, again resulting in an underdefined relation.

As it is, any modification to one of the constitutive laws does not canonically induce any modification to the other. Rather, the collection of mappings $\{C_\text{int},C_\text{ext},\phi_\text{i-e},\phi_\text{e-i}\}$ is an independent object. This is a more category-theoretic description of the simple truth that gravitational and internal gauge interactions are (apparently, and in the standard description) independent and define independent degrees of freedom.

It is possible to promote the constitutive mappings into dynamical objects in a very straightforward manner, but this would appear as just different field content, so there is no capacity for any different phenomenology either. A field is understood as a mapping from spacetime to some field space, so mapped again, the result is simply a field in a different space: for example, pointwise $\Phi(x)=C_\text{int}(B)\vert_x=*F\vert_x$. In another vein, let us assume that the constitutive mappings are linear, so it is possible to define their action on basis sections $\sigma_I$. As the bundles of differential forms are locally Euclidean, the mappings can be locally understood as functions $\mathbb{R}^k\to\mathbb{R}^l$, and can be written simply as
\begin{equation}
	C_\text{int}(B)=C_\text{int}(B^I_{\mu_1\ldots\mu_{n-2}} T_\mathfrak{g}^I\der x^{\mu_1}\wedge\ldots\wedge\der x^{\mu_{n-2}})=
	B^I_{\nu_1\ldots\nu_{n-2}}c_\text{int}^{IJ \nu_1\ldots\nu_{n-2}}{}_{\mu_1\ldots\mu_{n-2}}T_{\mathfrak{g}_\text{int}}^J \der x^{\mu_1}\wedge\ldots\wedge\der x^{\mu_{n-2}}.
\end{equation}
In practice, this introduces a $n-2$ form $c_\text{int}^{I \nu_1\ldots\nu_{n-2}}$ valued in the Lie algebra $\mathfrak{g}_\text{int}$ as an independent field. The case for $C_\text{ext},\phi_\text{i-e},\phi_\text{e-i}$ is analogous. In particular, for $\phi_\text{e-i}$, the situation is trivial, as
\begin{equation}
	\begin{aligned}
		\phi_{e-i}(*(e^a\wedge e^b))&=\frac{1}{2!(n-2)!}\epsilon^{ab}{}_{i_1\ldots i_{n-2}}\phi_{e-i\ ab}^{\phantom{e-i\ }I}{}^{i_1\ldots i_{n-2}}{}_{j_1\ldots j_{n-2}}T^I e^{j_1}\wedge\ldots\wedge e^{j_{n-2}}\\
		&=B_\text{int}\equiv *F_\text{int}.
	\end{aligned}
\end{equation}
That is, because $\epsilon^{ab}{}_{i_1\ldots i_{n-2}}$ is not dynamical, the result is simply a new differential form, which is assumed to be equal to the dual field strength.

Note that it is not even clear whether there ever can be a nontrivial gravitational constitutive law, because if the external, gravitational excitation cannot be interpreted in terms of the dual surface element $*(e^a\wedge e^b)$, the would-be-gravitational theory does not hold an interpretation in terms of Einstein-Cartan or metric-affine gravity. Therefore, it can instead be argued that the constitutive mappings $\{C_\text{int},C_\text{ext},\phi_\text{i-e},\phi_\text{e-i}\}$ should be trivial, in the sense that they constantly map to only one output element $\{B,*(e^a\wedge e^b),*F\}$ regardless of the input. Alternatively, it is straightforward to introduce a gravitational interpretation equation
\begin{equation}
	*'({e'}^a\wedge {e'}^b)=B^{ab}_\text{ext}[\phi_\text{all}],
\end{equation}
which should be solved for the new coframe ${e'}^a$, while $B^{ab}_\text{ext}[\phi_\text{all}]$ is a nontrivial expression of some field content $\phi_\text{all}$. The extent to which this is or is not possible determines the extent to which the nontrivial constitutive law can be interpreted in Einstein-Cartan gravitational terms. As a separate consistency requirement, the curvature content $R[\omega_\text{ext}]$ should not introduce additional degrees of freedom as well; that is, it should be determined by $B^{ab}_\text{ext}[\phi_\text{all}]$.

What this issue shows is not that any particular use of a special form of the gravitational excitation would be impossible or unwarranted, but just that there is no canonical method to accomplish this. Instead, the obvious way is to simply take any gauge-gravity BF-type unified theory, and (if possible and consistent) induce the specific form of the excitation directly onto the internal or external excitation.  For a unified model specifically, consider e.g.~the Plebanski unified model~\cite{Smolin:2009}, whose Lagrangian is
\begin{equation}
	S_\text{Pl.G}=\frac{1}{G}\int B^{A}\wedge R_A - \frac{1}{2}\phi_{AB}B^A\wedge B^B + \frac{g}{2}\phi^{AB}\phi_{AB}B^C\wedge B_C.
\end{equation}
Note $A,B,\ldots$ refer to generators of a general semisimple $G$, which contains the Euclidean $SO(4)$ as a subgroup. This theory is then to be resolved through the ansatz
\begin{equation}
	B^{ab}=e^a\wedge e^b + \gamma*(e^a\wedge e^b),\ \gamma\neq0,
\end{equation}
so it follows
\begin{equation}
	B^{ab}\wedge B_{ab}=\gamma\epsilon_{abcd}e^a\wedge e^b\wedge e^c\wedge e^d.
\end{equation}
The ansatz can be adapted to the ``Khronon'' Lorentz gauge theory~\eqref{eq:khronon} by simply moving to complexified space and appending constraints,
\begin{equation}
	S_{\text{Pl.G,}\tau}=S_\text{Pl.G} + \int X_{ab}\wedge(B^{ab} - \Der\tau^a\wedge\Der\tau^b - \gamma'*(\Der\tau^a\wedge\Der\tau^b)) + \int C_{ab}\wedge(R^{ab} - R^{+ab}).
\end{equation}
Note the restriction to the Lorentz-valued $B^{ab}$ rather than the total gauge group. When comparing with the ansatz, the surface element constraint would provide the additional implication
\begin{equation}
	e^a\wedge e^b + \gamma*(e^a\wedge e^b) = \Der\tau^a\wedge\Der\tau^b + \gamma'*(\Der\tau^a\wedge\Der\tau^b),
\end{equation}
where care should be taken that $\gamma'=\gamma$. But if so, as a particular solution
\begin{equation}
	e^a=\Der\tau^a,
\end{equation}
which is the geometric substance that the ``Khronon'' theory works around. In a sense, constraints of the type $B^{ab}\sim\Der\tau^a\wedge\Der\tau^b$ are a completion of this Lorentz gauge theory, because in the initial action~\eqref{eq:khronon} $e^a\sim\Der\tau^a$ is only implicit, and was only strictly enforced in the Hamiltonian analysis in~\cite{Zlosnik:2018qvg}. Neglecting the Immirzi parameter $\gamma$ for simplicity, note that the dust background is produced through the Lagrange multiplier, as the $\tau^a$ equations of motion imply
\begin{equation}
	\Der((*\lambda_{ab})\wedge\Der\tau^b)=0,
\end{equation}
and $\lambda_{ab}$ contributes to the Einstein equations. What has happened is that the degrees of freedom have been shuffled around, between Hamiltonian constraints, equations of motion and Lagrange multipliers, but the physical content is, of course, unchanged.

\section{Conclusion}

Let us reiterate the conclusion of the preceding analysis: there is no canonical way to implicate any particular constitutive law of one sector from the constitutive law of another sector, due to the simple fact that the interactions are independent. The contrary would have implied a very simple method to propose unified models from a particular premise. The constitutive diagram considered earlier is a different perspective into this issue. The correct method is now clear: the field theory content has to be analyzed in a case-by-case basis, entirely dependent on the intention of a particular model. Otherwise, the broken Lagrangian of any particular theory of gravity admits an equally obvious method to introduce any specific constitutive law into some other unified theory by introducing additional simplicity constraints.

The constraining of the $B$-field into its physical values has clear preference when considering what is external (spacetime related) and internal (matter content related). The issue is in executing the breaking. In BF theory alone, this requires some constraint structure, however care has to be taken to produce any meaningful physical addition, and not to simply redefine the field content. Unfortunately, the simplest $B$-potentials did not prove viable for ``spontaneous'' breaking in the sense of providing a non-unique solution structure to the auxiliary field, but this only directs to additional field content or a different representation of constraints. Plebanski theory~\cite{Plebanski:1977} is a famous example of nontrivial constraint structure, see also~\cite{Capovilla:1991kx,Capovilla:2001zi}, and Macdowell-Mansouri theory can similarly be put in BF form~\cite{Freidel:2005}. The analysis of the constraint structure can be pushed further, particularly in the quantum theory and path integrals, as it is also the premise for defining the parent action (path integral) for studying dualities.

Finally, the topology or manifold structure of the field theory could also be made more immediately dynamical, such as in the vein of~\cite{Borde:1994}. So far, $M$ has been simply postulated. The physical meaning of $M$ is to choose what is the model of signal transmission. That is, in manifold topology, this locally considers open balls in $\mathbb{R}^n$, with the metric defining the light cone. Note that a topology more directly causal, such as the Alexandrov or Zeeman-Göbel topology, is \emph{not} a manifold, and has subtler implications in terms of the Limit Curve Theorem or Gao-Wald theorem~\cite{Papadopoulos:2018uid,Papadopoulos:2021fex}. This would either suggest studying topological theories on nontrivial backgrounds, such as graphs~\cite{Antonsen:1994fka,Markopoulou:2007jf}, or retaining the manifold as a background and considering how the physical topology is embedded. Here, this is directly motivated from the observer-signal synchronization and spacetime construction model considered earlier, and would be interesting to propose in information-theoretic terms. As a particular physical system to consider, the genesis and elimination of spacetime singularities such as black holes could possibly obtain a different resolution.

\acknowledgments

The author would like to thank T. Koivisto and M. Hohmann for helpful discussions. This work was supported by the Estonian Research Council Grant No.~PRG2608 ``Space - Time - Matter'' and the Estonian Research Council CoE program with the Grant No.~TK202 ``Foundations of the Universe''.

\bibliography{references}

\end{document}